\documentclass[%
 reprint,
 amsmath,amssymb,
aps
pra,twocolumn,
]{revtex4-2}
\usepackage{braket}
\usepackage{graphicx}
\usepackage{bm}
\usepackage{float}
\usepackage{color}
\usepackage{soul}
\usepackage[caption = false]{subfig}
\usepackage{hyperref}
\usepackage{placeins}

\usepackage[normalem]{ulem}

\begin{document}

\title{Eliminating Thermal IR Background Noise by Imaging with Undetected Photons}
\author{Yue Ma$^1$}
\author{Nathan Gemmell$^1$}
\author{Emma Pearce$^1$}
\author{Rupert Oulton$^1$}
\author{Chris Phillips$^1$}

\affiliation{$^1$Blackett Laboratory, Imperial College London, London SW7 2AZ, United Kingdom\\
}

\begin{abstract}

Spectroscopy and imaging in the mid-infrared (2.5\,$\mu$m\,$\sim$\,$\lambda$\,$\sim$\,25\,$\mu$m) is bedevilled by the presence of a strong 300\,K thermal background at room temperature that makes IR detectors decades noisier than can be readily achieved in the visible. The technique of “imaging with undetected photons” (IUP) exploits the quantum correlations between entangled photon pairs to transfer image information from one spectral region to another, and here we show that it does so in a way that is immune to the thermal background. This means that IUP can be used to perform high speed photon counting measurements across the mid-IR, using uncooled visible detectors that are many times cheaper, faster, and more sensitive than their IR counterparts. 

\end{abstract}

\date{\today}

\maketitle

\section{introduction}

The so-called “fingerprint region” in the mid-infrared (mid-IR) spectrum (2.5 $\mu$m $\sim$ $\lambda$ $\sim$ 25 $\mu$m) is where many molecules absorb via bond-specific vibronic transitions. This makes it useful for chemical analysis via  vibrational spectroscopy in applications~\cite{salzer2014infrared} that span biochemical, environmental, pharmaceutical, and healthcare sectors. 

Recently, specialist spectroscopic imagers have been developed~\cite{amrania2009benchtop} to diagnose and track cancers via the chemical changes associated with the disease~\cite{amrania2018mid,amrania2016new}. However, measurements in this spectral region take place against a large and fluctuating background of thermal IR radiation that peaks at $\lambda\,\sim\,10\,\mu$m. The associated photon noise typically limits the achievable detectivity of even the best cooled detectors: the so-called ``background limited infrared photodetector" (BLIP) limit. Moreover, the $1/f$ noise character of its intensity fluctuations makes it challenging to eliminate in room temperature systems where the IR cameras are inherently slow as well as expensive.

As an example, a cooled $1\, \mathrm{mm}^2$ BLIP detector operating at the $\lambda\,\sim\,8\,\mu$m wavelength used to detect DNA in Ref.~\cite{amrania2009benchtop} would see $\sim\,11.7\,\mu$W of IR power in the $1176\,-\,1234\, \mathrm{cm}^{-1}$ measurement range, corresponding to a photon flux of $\sim\,5\,\times\,10^{14}\,\mathrm{sec}^{-1}$. Technical limitations mean that in fact, room temperature detectors are noisier than this by several decades.

The technique of  ``imaging with undetected photons" (IUP) in~\cite{lemos2014quantum} uses the quantum correlations between entangled photon pairs to transfer imageinformation from one photon to another with a different wavelength. It has the useful property that it allows an IR image to be registered with a detection system, such as a  CMOS silicon camera, at shorter wavelengths~\cite{buzas2020biological,paterova2020hyperspec,gilaberte2021video}, and the technique is being extended into the mid-IR~\cite{kviatkovsky2020microscopy,lindner21FTIR,paterova2022broadband,mukai2022fingerprint}. IUP is often realised using a nonlinear interferometer, such as we model here. At room temperature, a $\lambda\,\sim\,8\,\mu$m wavelength detector would see a black-body energy $\sim\,3.5\,\times 10^{15}$ times the energy seen by one detecting at $\lambda\,\sim\,1\,\mu$m. IUP therefore promises to extend the capability for high speed and high sensitivity photon counting measurements across the mid-IR for the first time, by sidestepping this large thermal background.

However, to realise this advantage, we must first show that the thermal background does not itself significantly affect the undetected photon imaging system. This requires showing that thermal states cannot also affect the information in the signal channel by ``seeding" the parametric down-conversion (PDC) process in the non-linear crystals in the interferometer (Fig.~\ref{fig:model}). 

PDC seeded by thermal states has been studied in a single crystal system~\cite{degiovanni2007intensity,degiovanni2009monitoring}. Separately, the operation of a nonlinear interferometer restricted to Gaussian states has been analysed~\cite{sparaciari2016gaussian}, but only with photon detection in both paths. More recently the photon measurement of only one path of a seeded nonlinear interferometer has been considered~\cite{florez2022enhanced}, but only with seeding by coherent states and number states.

Here we investigate the effect of thermal background radiation on a nonlinear interferometer where one path is seeded by a thermal state, while the photon number in the other path is measured. This corresponds to the realistic  scenario where the image information is extracted from the short wavelength channel, where the detectors are cheap and/or much more sensitive. We find that the quantity which corresponds to the IUP image quality, namely  the visibility of the interference fringes in the interferometer, is not affected by thermal seeding by near-infrared wavelengths ($\sim1\,\mu$m). For much longer wavelengths ($\sim8\,\mu$m), the thermal seeding only starts to impact the IUP image quality at black-body temperatures above $\sim750$\,K. In fact, if the whole interferometer system is held at the same temperature, we find that it is completely immune to thermal seeding effects up to extremely high temperatures. This implies that IUP imaging with undetected photons can offer a route to precision IR spectroscopy even in hostile environments where conventional approaches would be impossible.

\section{The model}

\begin{figure}[t]
\centering
\includegraphics[width=0.45\textwidth]{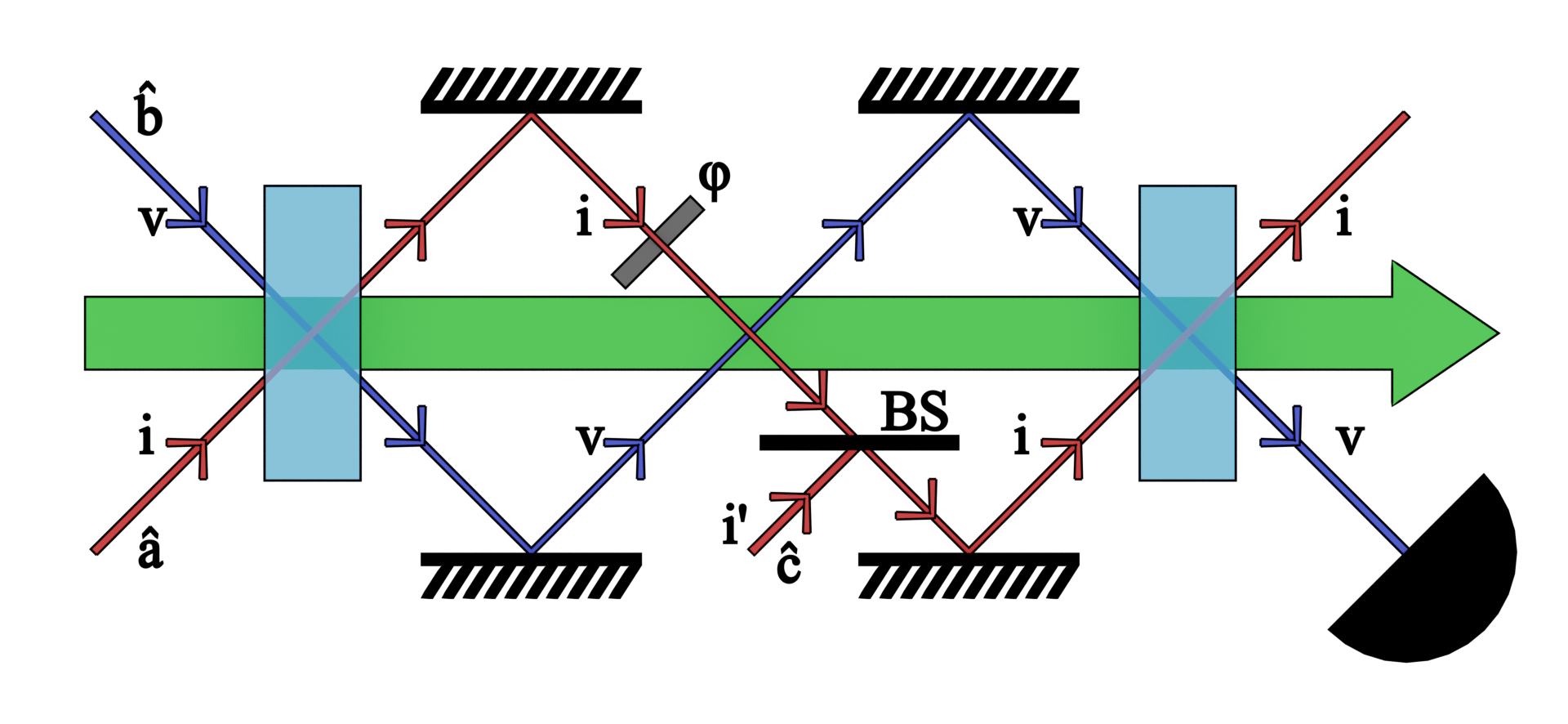}
\caption{Schematic of the nonlinear interferometer model. Two nonlinear crystals are pumped by the same continuous-wave laser. Their input modes are treated quantum mechanically. The photon annihilation operator $\hat{b}$ acts on mode v, which has a frequency in the visible range and is unseeded. The other input mode, i, has an IR frequency, and is acted on by photon annihilation operator $\hat{a}$; this mode may be seeded by a thermal state. The addition of the beam-splitter (BS) allows us to model the effect of inputting a thermal seed into the second crystal independently. This models the radiation emitted by objects within the interferometer. The two input ports of the beam-splitter are the infrared mode i and another mode i'. i' is also seeded by a thermal state, with a photon annihilation operator $\hat{c}$. A variable phase shifter ($\phi$) is included in the IR mode i, to allow us to model the interference fringe visibility. }\label{fig:model}
\end{figure}

PDC is an intrinsically pairwise process~\cite{degiovanni2007intensity,degiovanni2009monitoring}, so we need only perform an explicit  derivation for the case where each of the two input modes are monochromatic~\cite{zou1991induced,leonhardt2003quantum,grynberg2010introduction}. In the model [Fig.~\ref{fig:model}] two nonlinear crystals are pumped by a common continuous-wave laser, frequency $\omega_\mathrm{p}$, in the ``low-gain" regime (where the pump is not depleted) so the pump is unchanged by the PDC process and is approximated as being classical~\cite{grynberg2010introduction}. The two modes that are seeded into the first crystal and fed by the PDC process are labelled as paths i and v, having frequencies $\omega_\mathrm{i}$ and $\omega_\mathrm{v}$, and corresponding to the IR and visible ranges respectively. Energy conservation implies that $\omega_\mathrm{i}+\omega_\mathrm{v}=\omega_\mathrm{p}$.

The fringe visibility is modelled by varying the phase of a phase shifter present in the infrared mode between the two crystals. Also included is a variable beam-splitter (BS) in the IR mode before it enters the second crystal, which allows us to mix in an independent seeding state through path i', also at $\omega_\mathrm{i}$ with the output state of the first crystal.

With the beam-splitter set at $100\%$ transmission the IR input state to the second crystal contains only the IR output state from the first,  whereas for a $100\%$ reflection it contains only the additional i' thermal mode which is independent of the first crystal. This beam-splitter both acts an avenue by which we can model the insertion of thermal radiation, and also as a means of modelling an object whose transmissivity is to be probed by the system. The photon number in the visible mode is measured at the end of the nonlinear interferometer.

\section{Evolution of the state}

We use quantum mechanics to describe the infrared mode $\mathrm{i}$, the visible mode $\mathrm{v}$, and the additional mode $\mathrm{i'}$ for seeding the second crystal. The annihilation (creation) operator in each mode is denoted $\hat{a}$ ($\hat{a}^{\dagger}$), $\hat{b}$ ($\hat{b}^{\dagger}$), and $\hat{c}$ ($\hat{c}^{\dagger}$), respectively. We consider only thermal seeding of the IR modes, $\mathrm{i}$ and $\mathrm{i'}$. The initial state of the system is therefore the tensor product of three density matrices,
\begin{equation}
    \rho_{0}=\rho_{\mathrm{th,i}}\otimes|0\rangle_{\mathrm{v}}\langle0|\otimes\rho_{\mathrm{th,c}},
\end{equation}
where $\rho_{\mathrm{th,i}}$ is a thermal state for mode $\mathrm{i}$~\cite{loudon2000quantum,mandel1995optical},
\begin{equation}
    \rho_{\mathrm{th,i}}=\sum_{n=0}^{\infty}p_n|n\rangle_i\langle n|.\label{eq:th1}
\end{equation}
Here the probabilities $p_n$ satisfy the Bose-Einstein distribution,
\begin{equation}
    p_n=\frac{\exp[-n\hbar\omega_\mathrm{i}/k_{\mathrm{B}}T]}{\sum_{m=0}^{\infty}\exp[-m\hbar\omega_\mathrm{i}/k_{\mathrm{B}}T]}\label{eq:th2}
\end{equation}
where $k_{\mathrm{B}}$ is the Boltzmann constant, $T$ is the temperature, and $|n\rangle$ is a number state. The mean photon number, defined as $n_{\mathrm{th,i}}=\mathrm{Tr}(\rho_{\mathrm{th,i}}\hat{a}^{\dagger}\hat{a})$, can be found straightforwardly as
\begin{equation}
    n_{\mathrm{th,i}}=\frac{1}{\exp[\hbar\omega_{\mathrm{i}}/k_{\mathrm{B}}T]-1}.\label{eq:th3}
\end{equation}
The unseeded visible mode $\mathrm{v}$ is initially the vacuum state $|0\rangle_{\mathrm{v}}\langle0|$, and the thermal state $\rho_{\mathrm{th,c}}$ denotes the additional mode $\mathrm{i'}$. As is the case in Eqs.~\eqref{eq:th1}-\eqref{eq:th3}, $\rho_{\mathrm{th,c}}$ has a mean photon number
\begin{equation}
    n_{\mathrm{th,c}}=\frac{1}{\exp[\hbar\omega_{\mathrm{i}}/k_{\mathrm{B}}T']-1}.\label{eq:th4}
\end{equation}
Note that although modes $\mathrm{i'}$ and $\mathrm{i}$ have the same frequency, $\omega_{\mathrm{i}}$, the thermal states $\rho_{\mathrm{th,c}}$ and $\rho_{\mathrm{th,i}}$ correspond to temperatures $T$ and $T'$ which may differ from each other.

We describe the effect of each optical element in the Schr\"{o}dinger picture. After passing the first crystal, the state of the system evolves into
\begin{equation}
    \rho_1=\exp[i\xi(\hat{a}\hat{b}+\hat{a}^{\dagger}\hat{b}^{\dagger})]\rho_0\exp[-i\xi(\hat{a}\hat{b}+\hat{a}^{\dagger}\hat{b}^{\dagger})].
\end{equation}
The unitary operator $\exp[i\xi(\hat{a}\hat{b}+\hat{a}^{\dagger}\hat{b}^{\dagger})]$ is the two-mode squeezing operator~\cite{walls2008quantum}, describing the PDC process involving mode $\mathrm{i}$ and mode $\mathrm{v}$, as described by the transformation
\begin{equation}
    e^{i\xi(\hat{a}\hat{b}+\hat{a}^{\dagger}\hat{b}^{\dagger})}\hat{a} e^{-i\xi(\hat{a}\hat{b}+\hat{a}^{\dagger}\hat{b}^{\dagger})}=\cosh\xi\cdot\hat{a}-i\sinh\xi\cdot\hat{b}^{\dagger},
\end{equation}
where the parametric gain, $\xi$, is assumed to be in the low-gain regime ($\xi\ll 1$) and, without loss of generality, to be real and positive. After the infrared mode $\mathrm{i}$ passes through the variable phase shifter, the state of the system becomes
\begin{equation}
    \rho_2=\exp[i\varphi\hat{a}^{\dagger}\hat{a}]\rho_1\exp[-i\varphi\hat{a}^{\dagger}\hat{a}],
\end{equation}
where $\varphi$ is the phase shift,
\begin{equation}
    e^{-i\varphi\hat{a}^{\dagger}\hat{a}}\hat{a}^{\dagger}e^{i\varphi\hat{a}^{\dagger}\hat{a}}=a^{\dagger}e^{-i\varphi}.
\end{equation}
After the beam-splitter which mixes the state of mode $\mathrm{i}$ with the thermal state of mode $\mathrm{i'}$, the state of the system becomes
\begin{equation}
    \rho_3=\exp[i\kappa(\hat{a}\hat{c}^{\dagger}+\hat{a}^{\dagger}\hat{c})]\rho_2\exp[-i\kappa(\hat{a}\hat{c}^{\dagger}+\hat{a}^{\dagger}\hat{c})].
\end{equation}
Here the unitary operator $\exp[i\kappa(\hat{a}\hat{c}^{\dagger}+\hat{a}^{\dagger}\hat{c})]$ is the beam-splitter operator~\cite{walls2008quantum} where the transmissivity is parameterised by the variable $0\leq\kappa\leq\pi/2$. The transmissivity and reflectivity of the beam-splitter are $\cos^2\kappa$ and $\sin^2\kappa$, respectively~\cite{walls2008quantum}, as described by the unitary transformation
\begin{equation}
    e^{-i\kappa(\hat{a}\hat{c}^{\dagger}+\hat{a}^{\dagger}\hat{c})}\hat{a}^{\dagger}e^{i\kappa(\hat{a}\hat{c}^{\dagger}+\hat{a}^{\dagger}\hat{c})}=\cos\kappa\cdot\hat{a}^{\dagger}-i\sin\kappa\cdot\hat{c}^{\dagger}.
\end{equation}
Note that although the quantum states corresponding to the modes $\mathrm{i}$ and $\mathrm{i'}$ are mixed on the beam-splitter, their labels remain unchanged, namely, mode $\mathrm{i}$ still corresponds to the path that enters the second crystal. Finally, mode $\mathrm{i}$ and mode $\mathrm{v}$ pass the second crystal, resulting in the evolved state
\begin{equation}
    \rho_4=\exp[i\xi(\hat{a}\hat{b}+\hat{a}^{\dagger}\hat{b}^{\dagger})]\rho_3\exp[-i\xi(\hat{a}\hat{b}+\hat{a}^{\dagger}\hat{b}^{\dagger})].
\end{equation}
For simplicity, we assume that both crystals have the same gain parameter $\xi$.

\section{VISIBILITY OF THE INTERFERENCE FRINGES}\label{sec:visibility}

In practice, the best detectors are available in the visible, so we measure only the mean photon number in mode $\mathrm{v}$, given by $N_{\mathrm{v}}=\mathrm{Tr}(\rho_4\hat{b}^{\dagger}\hat{b})$ and calculated to be
\begin{align}
    N_{\mathrm{v}}=&(\cosh^2\xi-1)\times\nonumber\\
    &\Big((n_{\mathrm{th,i}}+1)\cosh^2\xi(\cos^2\kappa+1+2\cos\kappa\cos\varphi)\nonumber\\
    &+(n_{\mathrm{th,c}}+1)(1-\cos^2\kappa)\Big).\label{eq:Nv}
\end{align}

The interference fringe visibility is defined as
\begin{equation}
    V=\frac{N_{\mathrm{v,max}}-N_{\mathrm{v,min}}}{N_{\mathrm{v,max}}+N_{\mathrm{v,min}}}.
\end{equation}
Note that, although in general, we must find the  extremal values of $N_{\mathrm{v}}$ over all possible values of $\varphi$, in this case $N_{\mathrm{v,max}}$ and  $N_{\mathrm{v,min}}$  occur at $\varphi=0$ and  $\pi$ respectively. Thus the visibility is
\begin{align}
    &V=\nonumber\\
    &\frac{2(n_{\mathrm{th,i}}+1)\cosh^2\xi\cdot\cos\kappa}{(n_{\mathrm{th,i}}+1)\cosh^2\xi\cdot(1+\cos^2\kappa)+(n_{\mathrm{th,c}}+1)(1-\cos^2\kappa)}.\label{eq:V}
\end{align}

 Looking at Eq. (15) we note that the fringe visibility will be completely unaffected by the thermal background when the temperature is uniform throughout the interferometer i.e. when  $n_\mathrm{th,i}=n_\mathrm{th,c}$; the visibility can only start to deteriorate if the degree of thermal seeding differs between the two crystals. In fact, remarkably, if the outside temperature is the greater then the thermal background actually improves the fringe visibility.

\section{mean thermal photon number and black-body radiation}

Inspecting Eq.~\eqref{eq:Nv} one sees that the thermal photons always enter the calculation in the form ($n_{\mathrm{th,i}}+1$) and ($n_{\mathrm{th,c}}+1$), where the +1 contribution  results from the spontaneous nature of the PDC. Setting $n_{\mathrm{th,i}}=n_{\mathrm{th,c}}=0$, should reproduce the results of other models that ignore thermal seeding.

Using  Eqs.~\eqref{eq:th3} and~\eqref{eq:th4} we find that, at $300\ \mathrm{K}$, $n_{\mathrm{th,i}}$ and $n_{\mathrm{th,c}}$ evaluate as  $1.4 \times 10^{-21}$ and $2.5 \times 10^{-3}$ for $\omega_{\mathrm{i}}$  values corresponding to wavelengths of 1 $\mu$m  and 8 $\mu$m respectively. These are both $\ll1$, and therefore of negligible impact in Eq.~\eqref{eq:V}. We thus conclude that, even if the IUP approach is extended to wavelengths close to the peak of the 300 K black-body spectrum, the IUP image quality will be essentially immune to  the thermal IR background. 

At first this result may seem surprising, but it must be remembered that the spectrum of the black-body energy density takes the form~\cite{loudon2000quantum}
\begin{equation}
    \langle W(\omega,T) \rangle d\omega = \langle n(\omega,T) \rangle \cdot \hbar \omega \cdot g(\omega)d\omega.\label{eq:planck}
\end{equation}
The first term on the right is the photon number (as given by Eq.~\eqref{eq:th3}), the second is the energy per photon, and the third, $g(\omega)=\omega^2/\pi^2c^3$, is the density of plane-wave photon modes per unit energy, where each has a wavevector $|k|=\omega/c$. The wavelength of the black-body peak, as described by Wien’s displacement law, is largely determined by the form of the density of states; at 300\,K, it occurs at a wavelength where the mean number of photons per mode is actually quite small. 

In our scheme the PDC is an interaction between 3 photon modes where each interaction involves an IR mode, $\mathbf{k}_\mathrm{i}$, a visible one,  $\mathbf{k}_\mathrm{v}$, and a pump photon, $\mathbf{k}_\mathrm{p}$, satisfying the phase-matching condition $\mathbf{k}_\mathrm{p}=\mathbf{k}_\mathrm{i}+\mathbf{k}_\mathrm{v}$~\cite{grynberg2010introduction}. By its nature, the PDC process occurs even without seeding photons, and we can see that in all cases of practical interest here (i.e. at 300\,K) , the effect of thermal seeding is small enough to be ignored. As the argument above applies to each pairwise process involving $\mathbf{k}_\mathrm{i}$ and $\mathbf{k}_\mathrm{v}$, it applies to the whole process when all the participating modes $\mathbf{k}_\mathrm{i}$ are taken into account. In particular, experimental parameters such as the joint spectral density specific to each nonlinear crystal contribute to the weighted sum of different pairwise processes, not the thermal photon number within each process.

We note also that $n_{\mathrm{th}}\ll1$ holds even for optical elements that are usually regarded as strong thermal emitters. For example, a 3000\,K incandescent light bulb would only have  $n_{\mathrm{th}}\,\sim\,10^{-3}$ for 600\,nm wavelength modes~\cite{mandel1995optical}, i.e. the momentum modes of the radiation field are still mostly empty. 

Next we consider higher temperature environments, to estimate when thermal backgrounds will actually start to degrade the imaging, and whether or not it matters which of the two crystals is being seeded.

\section{longer wavelength and higher temperature}

\begin{figure}[t]
\centering
\includegraphics[width=0.48\textwidth]{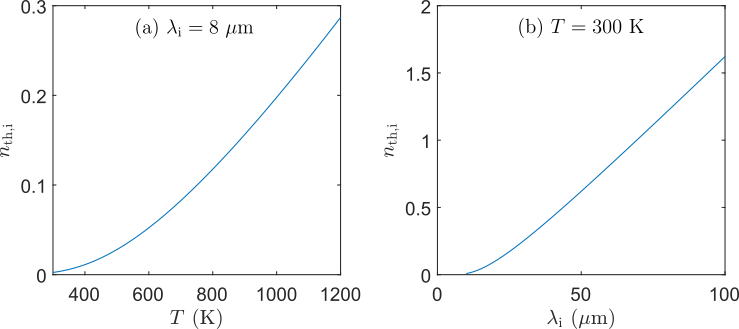}
\caption{The mean thermal photon number as a function of (a) temperature, fixing the wavelength as $8\ \mu$m. (b) wavelength, fixing the temperature as $300$ K.}\label{fig:number}
\end{figure}

\begin{figure}[t]
\centering
\includegraphics[width=0.48\textwidth]{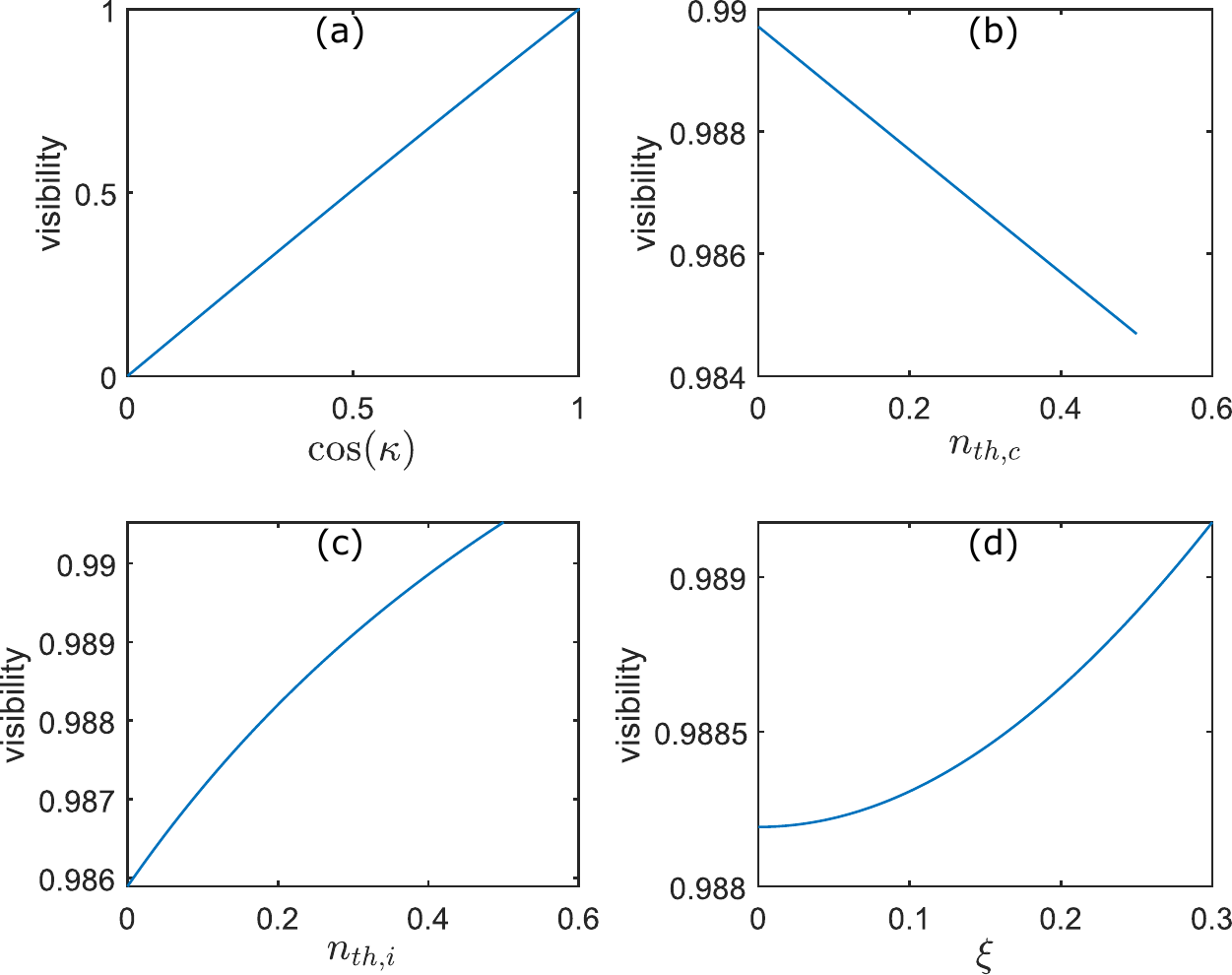}
\caption{Visibility of the nonlinear interferometer as a function of: (a) The square root of the beam-splitter transmissivity, $\cos\kappa$, where we have chosen $n_{\mathrm{th,i}}=0.2$, $n_{\mathrm{th,c}}=0.15$, $\xi=0.03$. (b) The mean photon number of the thermal seeding of the second crystal $n_{\mathrm{th,c}}$, where we have chosen $n_{\mathrm{th,i}}=0.2$, $\xi=0.03$, $\kappa=0.05\pi$. (c) The mean photon number of the thermal seeding of the first crystal $n_{\mathrm{th,i}}$, where we have chosen $n_{\mathrm{th,c}}=0.15$, $\xi=0.03$, $\kappa=0.05\pi$. (d) The squeezing parameter of the nonlinear crystals $\xi$, where we have chosen $n_{\mathrm{th,i}}=0.2$, $n_{\mathrm{th,c}}=0.15$, $\kappa=0.05\pi$.}\label{fig:visibility}
\end{figure}

As seen in Eq.~\eqref{eq:th3}, the  values of $n_{\mathrm{th,i}}$ and $n_{\mathrm{th,c}}$ will clearly increase if either the IR frequency decreases or the ambient temperature increases. For an IR  wavelength of $8\ \mu\mathrm{m}$, $n_{\mathrm{th}}\approx2.5\times10^{-3}$ at room temperature $T=300\,\mathrm{K}$, but increases to $\sim $0.1 by 750\,K, at which point one could imagine that the thermal IR background may start to degrade the image. However, even at this level, the attainable image quality using IUP is still likely be many decades better than what one can currently achieve with room temperature direct IR detectors at this wavelength. 

To be specific, in Fig.~\ref{fig:number} we show how the mean photon number $n_{\mathrm{th,i}}$ (or equivalently $n_{\mathrm{th,c}}$) changes as a function of temperature, and as a function of the considered wavelength [see Eq.~\eqref{eq:th3}]. For a wavelength of $8\ \mu$m, increasing the temperature will increase the mean thermal photon number, and from $T=750$ K, $n_{\mathrm{th,i}}$ starts to reach 0.1 and impact the PDC process. For a temperature of $300$ K,  $n_{\mathrm{th,i}}$ only becomes larger than 0.1 for wavelengths larger than $20\ \mu$m, 

Now we analyze how the visibility in Eq.~\eqref{eq:V} changes with respect to the parameters $\kappa$, $n_{\mathrm{th,c}}$, $n_{\mathrm{th,i}}$ and $\xi$.

First we examine how the visibility changes with the beam-splitter transmissivity. We define the square root of the transmissivity as $t=\cos\kappa$. It is easy to show that
\begin{equation}
    \frac{\partial V}{\partial t}>0.\label{eq:partial_t}
\end{equation}
That is to say that adding a thermal IR seed field at a point between the two crystals will always reduce the quality of the IUP image. This is because the IR output field from the first crystal has a phase that is linked to that of its visible output in a non-stochastic way. This relative phase information is partially destroyed, and IUP fringe visibility consequently reduced, when the thermal IR, with its entirely random phase, is added in.  When $t$ is reduced to 0, the fringe visibility also vanishes, as the IR entering the the second crystal is composed completely of random phase-independent thermal seeding. Both the maximum and the zero values of fringe visibility (corresponding to $t=1$ and 0 respectively), are independent of other parameters such as $n_{\mathrm{th,i}}$, $n_{\mathrm{th,c}}$ or $\xi$. In Fig.~\ref{fig:visibility}(a), we plot the visibility as a function of $\cos\kappa$, fixing the other parameters as $n_{\mathrm{th,i}}=0.2$, $n_{\mathrm{th,c}}=0.15$, $\xi=0.03$.

We next study how the visibility changes with the mean photon number of the thermal state to be seeded into the second crystal. Again we find
\begin{equation}
    \frac{\partial V}{\partial n_{\mathrm{th,c}}}<0.
\end{equation}
With similar reasoning as in the previous paragraph, the more the phase of the beam entering the second crystal is disrupted by the addition of the randomly-phased thermal radiation, the worse the effect on the IUP image. In Fig.~\ref{fig:visibility}(b), we plot the visibility as a function of $n_{\mathrm{th,c}}$, fixing the other parameters as $n_{\mathrm{th,i}}=0.2$, $\xi=0.03$, $\kappa=0.05\pi$. Note that even for $n_{\mathrm{th,c}}=0$, the visibility is less than 1. This is because we have set the  beam-splitter transmissivity to be less than unity, thereby mixing the output of the first crystal with the vacuum.  Since the vacuum does not have a well-defined phase, mixing it with the output of the first crystal before entering the second crystal also reduces the interference visibility.

Next we consider how the fringe visibility changes with the mean photon number of the thermal state seeding the first crystal. We have
\begin{equation}
    \frac{\partial V}{\partial n_{\mathrm{th,i}}}>0.\label{eq:partial_i}
\end{equation}
That is to say that even though increasing the mean photon number of the thermal state leads to a noisier seed, the visibility increases because this thermal state seeds the entire interferometer. Although the thermal state has a random phase, the relative phase between the down-converted IR and the visible field remains non-stochastic throughout the interferometer. A noisier thermal state also has a larger seeding power and creates more down-converted photon pairs. Despite mixing with the second seeding with a random phase, the overall number of photons carrying non-stochastic phase information is larger, resulting in a larger fringe visibility. In Fig.~\ref{fig:visibility}(c), we plot the visibility as a function of $n_{\mathrm{th,i}}$, fixing the other parameters as $n_{\mathrm{th,c}}=0.15$, $\xi=0.03$, $\kappa=0.05\pi$.

We can also look at how the visibility changes with the parametric gain $\xi$, which is
\begin{equation}
    \frac{\partial V}{\partial \xi}>0.
\end{equation}
A larger gain means more photon pairs are generated in the PDC. Similar to the analysis beneath Eq.~\eqref{eq:partial_i}, after the beam-splitter, the ratio of the number of coherent photons (whose phase relative to the visible path is well defined)versus photons with a random phase gets larger, thus increasing the visibility. In Fig.~\ref{fig:visibility}(d), we plot the visibility as a function of $\xi$, fixing the other parameters as $n_{\mathrm{th,i}}=0.2$, $n_{\mathrm{th,c}}=0.15$, $\kappa=0.05\pi$.

\section{Conclusion and discussion}

We consider a nonlinear interferometer setup where the idler is in the IR frequency range and the signal is in the visible frequency range. A variable phase shifter is present in the idler path, while only the photon number of the signal is measured. We have analytically shown that, if the idler mode is seeded by thermal states before entering both the first and the second crystal, the interference visibility depends on the mean photon number of the seeding thermal states, which is close to 0 for near-infrared and mid-infrared light at room temperature. This implies that the nonlinear interferometer is insensitive to thermal noise from the environment, even when there is a substantial background of IR  black-body radiation at the operating wavelength. This is in stark contrast to conventional photon detection schemes where photon noise from the thermal background radiation severely limits the system sensitivity. 

We further analyze the regime of longer wavelengths and higher temperatures, showing how the visibility responds differently to a change in the seeding of the first crystal or the second crystal. If the two crystals are subject to the same thermal background in the infrared mode, the seeding does not impact on the visibility. 

However, the visibility can actually be increased by isolating the second crystal from the thermal background noise that is injected from inside the interferometer, for example by cooling  some of its components. Also the visibility can be improved by injecting thermal radiation at the input mode i, e.g. by shining radiation from a hot object into the interferometer.

\acknowledgements  

We acknowledge funding from the UK National Quantum
Hub for Imaging (QUANTIC, No. EP/T00097X/1), an EPSRC
DTP, and the Royal Society (No. UF160475). Helpful discussions with Myungshik Kim and Martin McCall are gratefully acknowledged.


\begin{thebibliography}{22}%
\makeatletter
\providecommand \@ifxundefined [1]{%
 \@ifx{#1\undefined}
}%
\providecommand \@ifnum [1]{%
 \ifnum #1\expandafter \@firstoftwo
 \else \expandafter \@secondoftwo
 \fi
}%
\providecommand \@ifx [1]{%
 \ifx #1\expandafter \@firstoftwo
 \else \expandafter \@secondoftwo
 \fi
}%
\providecommand \natexlab [1]{#1}%
\providecommand \enquote  [1]{``#1''}%
\providecommand \bibnamefont  [1]{#1}%
\providecommand \bibfnamefont [1]{#1}%
\providecommand \citenamefont [1]{#1}%
\providecommand \href@noop [0]{\@secondoftwo}%
\providecommand \href [0]{\begingroup \@sanitize@url \@href}%
\providecommand \@href[1]{\@@startlink{#1}\@@href}%
\providecommand \@@href[1]{\endgroup#1\@@endlink}%
\providecommand \@sanitize@url [0]{\catcode `\\12\catcode `\$12\catcode
  `\&12\catcode `\#12\catcode `\^12\catcode `\_12\catcode `\%12\relax}%
\providecommand \@@startlink[1]{}%
\providecommand \@@endlink[0]{}%
\providecommand \url  [0]{\begingroup\@sanitize@url \@url }%
\providecommand \@url [1]{\endgroup\@href {#1}{\urlprefix }}%
\providecommand \urlprefix  [0]{URL }%
\providecommand \Eprint [0]{\href }%
\providecommand \doibase [0]{https://doi.org/}%
\providecommand \selectlanguage [0]{\@gobble}%
\providecommand \bibinfo  [0]{\@secondoftwo}%
\providecommand \bibfield  [0]{\@secondoftwo}%
\providecommand \translation [1]{[#1]}%
\providecommand \BibitemOpen [0]{}%
\providecommand \bibitemStop [0]{}%
\providecommand \bibitemNoStop [0]{.\EOS\space}%
\providecommand \EOS [0]{\spacefactor3000\relax}%
\providecommand \BibitemShut  [1]{\csname bibitem#1\endcsname}%
\let\auto@bib@innerbib\@empty
\bibitem [{\citenamefont {Salzer}\ and\ \citenamefont
  {Siesler}(2014)}]{salzer2014infrared}%
  \BibitemOpen
  \bibfield  {author} {\bibinfo {author} {\bibfnamefont {R.}~\bibnamefont
  {Salzer}}\ and\ \bibinfo {author} {\bibfnamefont {H.~W.}\ \bibnamefont
  {Siesler}},\ }\href@noop {} {\emph {\bibinfo {title} {Infrared and Raman
  Spectroscopic Imaging}}}\ (\bibinfo  {publisher} {John Wiley \& Sons},\
  \bibinfo {year} {2014})\BibitemShut {NoStop}%
\bibitem [{\citenamefont {Amrania}\ \emph {et~al.}(2009)\citenamefont
  {Amrania}, \citenamefont {McCrow},\ and\ \citenamefont
  {Phillips}}]{amrania2009benchtop}%
  \BibitemOpen
  \bibfield  {author} {\bibinfo {author} {\bibfnamefont {H.}~\bibnamefont
  {Amrania}}, \bibinfo {author} {\bibfnamefont {A.}~\bibnamefont {McCrow}},\
  and\ \bibinfo {author} {\bibfnamefont {C.}~\bibnamefont {Phillips}},\
  }\bibfield  {title} {\bibinfo {title} {A benchtop, ultrafast infrared
  spectroscopic imaging system for biomedical applications},\ }\href@noop {}
  {\bibfield  {journal} {\bibinfo  {journal} {Review of Scientific
  Instruments}\ }\textbf {\bibinfo {volume} {80}},\ \bibinfo {pages} {123702}
  (\bibinfo {year} {2009})}\BibitemShut {NoStop}%
\bibitem [{\citenamefont {Amrania}\ \emph {et~al.}(2018)\citenamefont
  {Amrania}, \citenamefont {Woodley-Barker}, \citenamefont {Goddard},
  \citenamefont {Rosales}, \citenamefont {Shousha}, \citenamefont {Thomas},
  \citenamefont {McFarlane}, \citenamefont {Sroya}, \citenamefont
  {Wilhelm-Benartzi}, \citenamefont {Cocks} \emph {et~al.}}]{amrania2018mid}%
  \BibitemOpen
  \bibfield  {author} {\bibinfo {author} {\bibfnamefont {H.}~\bibnamefont
  {Amrania}}, \bibinfo {author} {\bibfnamefont {L.}~\bibnamefont
  {Woodley-Barker}}, \bibinfo {author} {\bibfnamefont {K.}~\bibnamefont
  {Goddard}}, \bibinfo {author} {\bibfnamefont {B.}~\bibnamefont {Rosales}},
  \bibinfo {author} {\bibfnamefont {S.}~\bibnamefont {Shousha}}, \bibinfo
  {author} {\bibfnamefont {G.}~\bibnamefont {Thomas}}, \bibinfo {author}
  {\bibfnamefont {T.}~\bibnamefont {McFarlane}}, \bibinfo {author}
  {\bibfnamefont {M.}~\bibnamefont {Sroya}}, \bibinfo {author} {\bibfnamefont
  {C.}~\bibnamefont {Wilhelm-Benartzi}}, \bibinfo {author} {\bibfnamefont
  {K.}~\bibnamefont {Cocks}}, \emph {et~al.},\ }\bibfield  {title} {\bibinfo
  {title} {Mid-infrared imaging in breast cancer tissue: an objective measure
  of grading breast cancer biopsies},\ }\href@noop {} {\bibfield  {journal}
  {\bibinfo  {journal} {Convergent Science Physical Oncology}\ }\textbf
  {\bibinfo {volume} {4}},\ \bibinfo {pages} {025001} (\bibinfo {year}
  {2018})}\BibitemShut {NoStop}%
\bibitem [{\citenamefont {Amrania}\ \emph {et~al.}(2016)\citenamefont
  {Amrania}, \citenamefont {Drummond}, \citenamefont {Coombes}, \citenamefont
  {Shousha}, \citenamefont {Woodley-Barker}, \citenamefont {Weir},
  \citenamefont {Hart}, \citenamefont {Carter},\ and\ \citenamefont
  {Phillips}}]{amrania2016new}%
  \BibitemOpen
  \bibfield  {author} {\bibinfo {author} {\bibfnamefont {H.}~\bibnamefont
  {Amrania}}, \bibinfo {author} {\bibfnamefont {L.}~\bibnamefont {Drummond}},
  \bibinfo {author} {\bibfnamefont {R.}~\bibnamefont {Coombes}}, \bibinfo
  {author} {\bibfnamefont {S.}~\bibnamefont {Shousha}}, \bibinfo {author}
  {\bibfnamefont {L.}~\bibnamefont {Woodley-Barker}}, \bibinfo {author}
  {\bibfnamefont {K.}~\bibnamefont {Weir}}, \bibinfo {author} {\bibfnamefont
  {W.}~\bibnamefont {Hart}}, \bibinfo {author} {\bibfnamefont {I.}~\bibnamefont
  {Carter}},\ and\ \bibinfo {author} {\bibfnamefont {C.}~\bibnamefont
  {Phillips}},\ }\bibfield  {title} {\bibinfo {title} {New {IR} imaging
  modalities for cancer detection and for intra-cell chemical mapping with a
  sub-diffraction mid-{IR} s-{SNOM}},\ }\href@noop {} {\bibfield  {journal}
  {\bibinfo  {journal} {Faraday Discussions}\ }\textbf {\bibinfo {volume}
  {187}},\ \bibinfo {pages} {539} (\bibinfo {year} {2016})}\BibitemShut
  {NoStop}%
\bibitem [{\citenamefont {Lemos}\ \emph {et~al.}(2014)\citenamefont {Lemos},
  \citenamefont {Borish}, \citenamefont {Cole}, \citenamefont {Ramelow},
  \citenamefont {Lapkiewicz},\ and\ \citenamefont
  {Zeilinger}}]{lemos2014quantum}%
  \BibitemOpen
  \bibfield  {author} {\bibinfo {author} {\bibfnamefont {G.~B.}\ \bibnamefont
  {Lemos}}, \bibinfo {author} {\bibfnamefont {V.}~\bibnamefont {Borish}},
  \bibinfo {author} {\bibfnamefont {G.~D.}\ \bibnamefont {Cole}}, \bibinfo
  {author} {\bibfnamefont {S.}~\bibnamefont {Ramelow}}, \bibinfo {author}
  {\bibfnamefont {R.}~\bibnamefont {Lapkiewicz}},\ and\ \bibinfo {author}
  {\bibfnamefont {A.}~\bibnamefont {Zeilinger}},\ }\bibfield  {title} {\bibinfo
  {title} {Quantum imaging with undetected photons},\ }\href@noop {} {\bibfield
   {journal} {\bibinfo  {journal} {Nature}\ }\textbf {\bibinfo {volume}
  {512}},\ \bibinfo {pages} {409} (\bibinfo {year} {2014})}\BibitemShut
  {NoStop}%
\bibitem [{\citenamefont {B{\'u}z{\'a}s}\ \emph {et~al.}(2020)\citenamefont
  {B{\'u}z{\'a}s}, \citenamefont {Wolff}, \citenamefont {Benedict},
  \citenamefont {Ormos},\ and\ \citenamefont {D{\'e}r}}]{buzas2020biological}%
  \BibitemOpen
  \bibfield  {author} {\bibinfo {author} {\bibfnamefont {A.}~\bibnamefont
  {B{\'u}z{\'a}s}}, \bibinfo {author} {\bibfnamefont {E.~K.}\ \bibnamefont
  {Wolff}}, \bibinfo {author} {\bibfnamefont {M.~G.}\ \bibnamefont {Benedict}},
  \bibinfo {author} {\bibfnamefont {P.}~\bibnamefont {Ormos}},\ and\ \bibinfo
  {author} {\bibfnamefont {A.}~\bibnamefont {D{\'e}r}},\ }\bibfield  {title}
  {\bibinfo {title} {Biological microscopy with undetected photons},\
  }\href@noop {} {\bibfield  {journal} {\bibinfo  {journal} {IEEE Access}\
  }\textbf {\bibinfo {volume} {8}},\ \bibinfo {pages} {107539} (\bibinfo {year}
  {2020})}\BibitemShut {NoStop}%
\bibitem [{\citenamefont {Paterova}\ \emph {et~al.}(2020)\citenamefont
  {Paterova}, \citenamefont {Maniam}, \citenamefont {Yang}, \citenamefont
  {Grenci},\ and\ \citenamefont {Krivitsky}}]{paterova2020hyperspec}%
  \BibitemOpen
  \bibfield  {author} {\bibinfo {author} {\bibfnamefont {A.~V.}\ \bibnamefont
  {Paterova}}, \bibinfo {author} {\bibfnamefont {S.~M.}\ \bibnamefont
  {Maniam}}, \bibinfo {author} {\bibfnamefont {H.}~\bibnamefont {Yang}},
  \bibinfo {author} {\bibfnamefont {G.}~\bibnamefont {Grenci}},\ and\ \bibinfo
  {author} {\bibfnamefont {L.~A.}\ \bibnamefont {Krivitsky}},\ }\bibfield
  {title} {\bibinfo {title} {Hyperspectral infrared microscopy with visible
  light},\ }\href@noop {} {\bibfield  {journal} {\bibinfo  {journal} {Science
  Advances}\ }\textbf {\bibinfo {volume} {6}} (\bibinfo {year}
  {2020})}\BibitemShut {NoStop}%
\bibitem [{\citenamefont {Gilaberte~Basset}\ \emph {et~al.}(2021)\citenamefont
  {Gilaberte~Basset}, \citenamefont {Hochrainer}, \citenamefont {T{\"o}pfer},
  \citenamefont {Riexinger}, \citenamefont {Bickert}, \citenamefont
  {Le{\'o}n-Torres}, \citenamefont {Steinlechner},\ and\ \citenamefont
  {Gr{\"a}fe}}]{gilaberte2021video}%
  \BibitemOpen
  \bibfield  {author} {\bibinfo {author} {\bibfnamefont {M.}~\bibnamefont
  {Gilaberte~Basset}}, \bibinfo {author} {\bibfnamefont {A.}~\bibnamefont
  {Hochrainer}}, \bibinfo {author} {\bibfnamefont {S.}~\bibnamefont
  {T{\"o}pfer}}, \bibinfo {author} {\bibfnamefont {F.}~\bibnamefont
  {Riexinger}}, \bibinfo {author} {\bibfnamefont {P.}~\bibnamefont {Bickert}},
  \bibinfo {author} {\bibfnamefont {J.~R.}\ \bibnamefont {Le{\'o}n-Torres}},
  \bibinfo {author} {\bibfnamefont {F.}~\bibnamefont {Steinlechner}},\ and\
  \bibinfo {author} {\bibfnamefont {M.}~\bibnamefont {Gr{\"a}fe}},\ }\bibfield
  {title} {\bibinfo {title} {Video-rate imaging with undetected photons},\
  }\href@noop {} {\bibfield  {journal} {\bibinfo  {journal} {Laser \& Photonics
  Reviews}\ }\textbf {\bibinfo {volume} {15}},\ \bibinfo {pages} {2000327}
  (\bibinfo {year} {2021})}\BibitemShut {NoStop}%
\bibitem [{\citenamefont {Kviatkovsky}\ \emph {et~al.}(2020)\citenamefont
  {Kviatkovsky}, \citenamefont {Chrzanowski}, \citenamefont {Avery},
  \citenamefont {Bartolomaeus},\ and\ \citenamefont
  {Ramelow}}]{kviatkovsky2020microscopy}%
  \BibitemOpen
  \bibfield  {author} {\bibinfo {author} {\bibfnamefont {I.}~\bibnamefont
  {Kviatkovsky}}, \bibinfo {author} {\bibfnamefont {H.~M.}\ \bibnamefont
  {Chrzanowski}}, \bibinfo {author} {\bibfnamefont {E.~G.}\ \bibnamefont
  {Avery}}, \bibinfo {author} {\bibfnamefont {H.}~\bibnamefont
  {Bartolomaeus}},\ and\ \bibinfo {author} {\bibfnamefont {S.}~\bibnamefont
  {Ramelow}},\ }\bibfield  {title} {\bibinfo {title} {Microscopy with
  undetected photons in the mid-infrared},\ }\href@noop {} {\bibfield
  {journal} {\bibinfo  {journal} {Science Advances}\ }\textbf {\bibinfo
  {volume} {6}},\ \bibinfo {pages} {eabd0264} (\bibinfo {year}
  {2020})}\BibitemShut {NoStop}%
\bibitem [{\citenamefont {Lindner}\ \emph {et~al.}(2021)\citenamefont
  {Lindner}, \citenamefont {Kunz}, \citenamefont {Herr}, \citenamefont {Wolf},
  \citenamefont {Kie{\ss}ling},\ and\ \citenamefont
  {K\"{u}hnemann}}]{lindner21FTIR}%
  \BibitemOpen
  \bibfield  {author} {\bibinfo {author} {\bibfnamefont {C.}~\bibnamefont
  {Lindner}}, \bibinfo {author} {\bibfnamefont {J.}~\bibnamefont {Kunz}},
  \bibinfo {author} {\bibfnamefont {S.~J.}\ \bibnamefont {Herr}}, \bibinfo
  {author} {\bibfnamefont {S.}~\bibnamefont {Wolf}}, \bibinfo {author}
  {\bibfnamefont {J.}~\bibnamefont {Kie{\ss}ling}},\ and\ \bibinfo {author}
  {\bibfnamefont {F.}~\bibnamefont {K\"{u}hnemann}},\ }\bibfield  {title}
  {\bibinfo {title} {Nonlinear interferometer for {Fourier-transform}
  mid-infrared gas spectroscopy using near-infrared detection},\ }\href
  {https://doi.org/10.1364/OE.415365} {\bibfield  {journal} {\bibinfo
  {journal} {Optics Express}\ }\textbf {\bibinfo {volume} {29}},\ \bibinfo
  {pages} {4035} (\bibinfo {year} {2021})}\BibitemShut {NoStop}%
\bibitem [{\citenamefont {Paterova}\ \emph {et~al.}(2022)\citenamefont
  {Paterova}, \citenamefont {Toa}, \citenamefont {Yang},\ and\ \citenamefont
  {Krivitsky}}]{paterova2022broadband}%
  \BibitemOpen
  \bibfield  {author} {\bibinfo {author} {\bibfnamefont {A.~V.}\ \bibnamefont
  {Paterova}}, \bibinfo {author} {\bibfnamefont {Z.~S.}\ \bibnamefont {Toa}},
  \bibinfo {author} {\bibfnamefont {H.}~\bibnamefont {Yang}},\ and\ \bibinfo
  {author} {\bibfnamefont {L.~A.}\ \bibnamefont {Krivitsky}},\ }\bibfield
  {title} {\bibinfo {title} {Broadband quantum spectroscopy at the fingerprint
  mid-infrared region},\ }\href@noop {} {\bibfield  {journal} {\bibinfo
  {journal} {ACS Photonics}\ } (\bibinfo {year} {2022})}\BibitemShut {NoStop}%
\bibitem [{\citenamefont {Mukai}\ \emph {et~al.}(2022)\citenamefont {Mukai},
  \citenamefont {Okamoto},\ and\ \citenamefont
  {Takeuchi}}]{mukai2022fingerprint}%
  \BibitemOpen
  \bibfield  {author} {\bibinfo {author} {\bibfnamefont {Y.}~\bibnamefont
  {Mukai}}, \bibinfo {author} {\bibfnamefont {R.}~\bibnamefont {Okamoto}},\
  and\ \bibinfo {author} {\bibfnamefont {S.}~\bibnamefont {Takeuchi}},\
  }\bibfield  {title} {\bibinfo {title} {Quantum {Fourier-transform} infrared
  spectroscopy in the fingerprint region},\ }\href
  {https://doi.org/10.1364/OE.455718} {\bibfield  {journal} {\bibinfo
  {journal} {Opt. Express}\ }\textbf {\bibinfo {volume} {30}},\ \bibinfo
  {pages} {22624} (\bibinfo {year} {2022})}\BibitemShut {NoStop}%
\bibitem [{\citenamefont {Degiovanni}\ \emph {et~al.}(2007)\citenamefont
  {Degiovanni}, \citenamefont {Bondani}, \citenamefont {Puddu}, \citenamefont
  {Andreoni},\ and\ \citenamefont {Paris}}]{degiovanni2007intensity}%
  \BibitemOpen
  \bibfield  {author} {\bibinfo {author} {\bibfnamefont {I.~P.}\ \bibnamefont
  {Degiovanni}}, \bibinfo {author} {\bibfnamefont {M.}~\bibnamefont {Bondani}},
  \bibinfo {author} {\bibfnamefont {E.}~\bibnamefont {Puddu}}, \bibinfo
  {author} {\bibfnamefont {A.}~\bibnamefont {Andreoni}},\ and\ \bibinfo
  {author} {\bibfnamefont {M.~G.}\ \bibnamefont {Paris}},\ }\bibfield  {title}
  {\bibinfo {title} {Intensity correlations, entanglement properties, and ghost
  imaging in multimode thermal-seeded parametric down-conversion: Theory},\
  }\href@noop {} {\bibfield  {journal} {\bibinfo  {journal} {Physical Review
  A}\ }\textbf {\bibinfo {volume} {76}},\ \bibinfo {pages} {062309} (\bibinfo
  {year} {2007})}\BibitemShut {NoStop}%
\bibitem [{\citenamefont {Degiovanni}\ \emph {et~al.}(2009)\citenamefont
  {Degiovanni}, \citenamefont {Genovese}, \citenamefont {Schettini},
  \citenamefont {Bondani}, \citenamefont {Andreoni},\ and\ \citenamefont
  {Paris}}]{degiovanni2009monitoring}%
  \BibitemOpen
  \bibfield  {author} {\bibinfo {author} {\bibfnamefont {I.}~\bibnamefont
  {Degiovanni}}, \bibinfo {author} {\bibfnamefont {M.}~\bibnamefont
  {Genovese}}, \bibinfo {author} {\bibfnamefont {V.}~\bibnamefont {Schettini}},
  \bibinfo {author} {\bibfnamefont {M.}~\bibnamefont {Bondani}}, \bibinfo
  {author} {\bibfnamefont {A.}~\bibnamefont {Andreoni}},\ and\ \bibinfo
  {author} {\bibfnamefont {M.}~\bibnamefont {Paris}},\ }\bibfield  {title}
  {\bibinfo {title} {Monitoring the quantum-classical transition in thermally
  seeded parametric down-conversion by intensity measurements},\ }\href@noop {}
  {\bibfield  {journal} {\bibinfo  {journal} {Physical Review A}\ }\textbf
  {\bibinfo {volume} {79}},\ \bibinfo {pages} {063836} (\bibinfo {year}
  {2009})}\BibitemShut {NoStop}%
\bibitem [{\citenamefont {Sparaciari}\ \emph {et~al.}(2016)\citenamefont
  {Sparaciari}, \citenamefont {Olivares},\ and\ \citenamefont
  {Paris}}]{sparaciari2016gaussian}%
  \BibitemOpen
  \bibfield  {author} {\bibinfo {author} {\bibfnamefont {C.}~\bibnamefont
  {Sparaciari}}, \bibinfo {author} {\bibfnamefont {S.}~\bibnamefont
  {Olivares}},\ and\ \bibinfo {author} {\bibfnamefont {M.~G.}\ \bibnamefont
  {Paris}},\ }\bibfield  {title} {\bibinfo {title} {Gaussian-state
  interferometry with passive and active elements},\ }\href@noop {} {\bibfield
  {journal} {\bibinfo  {journal} {Physical Review A}\ }\textbf {\bibinfo
  {volume} {93}},\ \bibinfo {pages} {023810} (\bibinfo {year}
  {2016})}\BibitemShut {NoStop}%
\bibitem [{\citenamefont {Fl{\'o}rez}\ \emph {et~al.}(2022)\citenamefont
  {Fl{\'o}rez}, \citenamefont {Pearce}, \citenamefont {Gemmell}, \citenamefont
  {Ma}, \citenamefont {Bressanini}, \citenamefont {Phillips}, \citenamefont
  {Oulton},\ and\ \citenamefont {Clark}}]{florez2022enhanced}%
  \BibitemOpen
  \bibfield  {author} {\bibinfo {author} {\bibfnamefont {J.}~\bibnamefont
  {Fl{\'o}rez}}, \bibinfo {author} {\bibfnamefont {E.}~\bibnamefont {Pearce}},
  \bibinfo {author} {\bibfnamefont {N.~R.}\ \bibnamefont {Gemmell}}, \bibinfo
  {author} {\bibfnamefont {Y.}~\bibnamefont {Ma}}, \bibinfo {author}
  {\bibfnamefont {G.}~\bibnamefont {Bressanini}}, \bibinfo {author}
  {\bibfnamefont {C.~C.}\ \bibnamefont {Phillips}}, \bibinfo {author}
  {\bibfnamefont {R.~F.}\ \bibnamefont {Oulton}},\ and\ \bibinfo {author}
  {\bibfnamefont {A.~S.}\ \bibnamefont {Clark}},\ }\bibfield  {title} {\bibinfo
  {title} {Enhanced nonlinear interferometry via seeding},\ }\href@noop {}
  {\bibfield  {journal} {\bibinfo  {journal} {arXiv preprint arXiv:2209.06749}\
  } (\bibinfo {year} {2022})}\BibitemShut {NoStop}%
\bibitem [{\citenamefont {Zou}\ \emph {et~al.}(1991)\citenamefont {Zou},
  \citenamefont {Wang},\ and\ \citenamefont {Mandel}}]{zou1991induced}%
  \BibitemOpen
  \bibfield  {author} {\bibinfo {author} {\bibfnamefont {X.}~\bibnamefont
  {Zou}}, \bibinfo {author} {\bibfnamefont {L.~J.}\ \bibnamefont {Wang}},\ and\
  \bibinfo {author} {\bibfnamefont {L.}~\bibnamefont {Mandel}},\ }\bibfield
  {title} {\bibinfo {title} {Induced coherence and indistinguishability in
  optical interference},\ }\href@noop {} {\bibfield  {journal} {\bibinfo
  {journal} {Physical review letters}\ }\textbf {\bibinfo {volume} {67}},\
  \bibinfo {pages} {318} (\bibinfo {year} {1991})}\BibitemShut {NoStop}%
\bibitem [{\citenamefont {Leonhardt}(2003)}]{leonhardt2003quantum}%
  \BibitemOpen
  \bibfield  {author} {\bibinfo {author} {\bibfnamefont {U.}~\bibnamefont
  {Leonhardt}},\ }\bibfield  {title} {\bibinfo {title} {Quantum physics of
  simple optical instruments},\ }\href@noop {} {\bibfield  {journal} {\bibinfo
  {journal} {Reports on Progress in Physics}\ }\textbf {\bibinfo {volume}
  {66}},\ \bibinfo {pages} {1207} (\bibinfo {year} {2003})}\BibitemShut
  {NoStop}%
\bibitem [{\citenamefont {Grynberg}\ \emph {et~al.}(2010)\citenamefont
  {Grynberg}, \citenamefont {Aspect},\ and\ \citenamefont
  {Fabre}}]{grynberg2010introduction}%
  \BibitemOpen
  \bibfield  {author} {\bibinfo {author} {\bibfnamefont {G.}~\bibnamefont
  {Grynberg}}, \bibinfo {author} {\bibfnamefont {A.}~\bibnamefont {Aspect}},\
  and\ \bibinfo {author} {\bibfnamefont {C.}~\bibnamefont {Fabre}},\
  }\href@noop {} {\emph {\bibinfo {title} {Introduction to Quantum Optics: From
  the Semi-classical Approach to Quantized Light}}}\ (\bibinfo  {publisher}
  {Cambridge university press},\ \bibinfo {year} {2010})\BibitemShut {NoStop}%
\bibitem [{\citenamefont {Loudon}(2000)}]{loudon2000quantum}%
  \BibitemOpen
  \bibfield  {author} {\bibinfo {author} {\bibfnamefont {R.}~\bibnamefont
  {Loudon}},\ }\href@noop {} {\emph {\bibinfo {title} {The Quantum Theory of
  Light}}}\ (\bibinfo  {publisher} {OUP Oxford},\ \bibinfo {year}
  {2000})\BibitemShut {NoStop}%
\bibitem [{\citenamefont {Mandel}\ and\ \citenamefont
  {Wolf}(1995)}]{mandel1995optical}%
  \BibitemOpen
  \bibfield  {author} {\bibinfo {author} {\bibfnamefont {L.}~\bibnamefont
  {Mandel}}\ and\ \bibinfo {author} {\bibfnamefont {E.}~\bibnamefont {Wolf}},\
  }\href@noop {} {\emph {\bibinfo {title} {Optical Coherence and Quantum
  Optics}}}\ (\bibinfo  {publisher} {Cambridge university press},\ \bibinfo
  {year} {1995})\BibitemShut {NoStop}%
\bibitem [{\citenamefont {Walls}\ and\ \citenamefont
  {Milburn}(2008)}]{walls2008quantum}%
  \BibitemOpen
  \bibfield  {author} {\bibinfo {author} {\bibfnamefont {D.~F.}\ \bibnamefont
  {Walls}}\ and\ \bibinfo {author} {\bibfnamefont {G.~J.}\ \bibnamefont
  {Milburn}},\ }\href@noop {} {\emph {\bibinfo {title} {Quantum Optics}}}\
  (\bibinfo  {publisher} {Springer Berlin, Heidelberg},\ \bibinfo {year}
  {2008})\BibitemShut {NoStop}%
\end{thebibliography}
\end{document}